\documentclass[%
reprint,
superscriptaddress,
 amsmath,amssymb,
 aps,
prb,
]{revtex4-1}

\usepackage{comment}
\usepackage{physics}
\usepackage[normalem]{ulem}
\usepackage{physics}
\usepackage{xparse}

\newcommand{\mbf}[1]{\ensuremath{\mathbf{#1}}}

\NewDocumentCommand{\rep}{s d<| d|>}{%
\IfBooleanTF{#1}{
   \IfValueTF{#2}{
       \IfValueTF{#3}{\braket{#2}{#3}}{\bra{#2}}
       }{
       \IfValueTF{#3}{\ket{#3}}{}
       }
   }{
   \IfValueTF{#2}{
       \IfValueTF{#3}{\braket*{#2}{#3}}{\bra*{#2}}
       }{
       \IfValueTF{#3}{\ket*{#3}}{}
       }
   }
}

\NewDocumentCommand{\rbra}{sm}{\IfBooleanTF{#1}{\rep*<#2|}{\rep<#2|}}
\NewDocumentCommand{\rket}{sm}{\IfBooleanTF{#1}{\rep*|#2>}{\rep|#2>}}
\NewDocumentCommand{\rbraket}{smm}{\IfBooleanTF{#1}{\rep*<#2||#3>}{\rep<#2||#3>}}

\NewDocumentCommand{\field}{o m e{_} e{^} o e{_} e{^}}{
\IfValueTF{#5}{\overline{
  #2\IfValueT{#3}{_#3}\IfValueT{#4}{^{\otimes #4}} %
  \otimes 
  #5\IfValueT{#6}{_#6}\IfValueT{#7}{^{\otimes #7}} %
  \IfValueT{#1}{;#1}
}}{
  \IfValueTF{#4}{\overline{
     #2\IfValueT{#3}{_#3}\IfValueT{#4}{^{\otimes #4}}
     \IfValueT{#1}{;#1}
  }}
  {#2\IfValueT{#3}{_#3}}
}
}

\NewDocumentCommand{\frho}{o e{_} e{^}}{
\field[#1]{\rho}_{#2}^{#3}
}

\newcommand{\e}{a}  %

\newcommand{\br}{\mbf{r}}
\newcommand{\bx}{\mbf{x}}
\newcommand{\bxhat}{\hat{\mbf{x}}}

\NewDocumentCommand{\ex}{e_}{
\IfValueTF{#1}{\e_{#1}\bx_{#1}}{\e\bx}
}  %

\NewDocumentCommand{\lm}{e_}{
\IfValueTF{#1}{l_{#1}m_{#1}}{lm}
}
\NewDocumentCommand{\nlm}{e_}{
\IfValueTF{#1}{n_{#1}\lm_{#1}}{n\lm}
}
\NewDocumentCommand{\enlm}{e_}{
\IfValueTF{#1}{\e_{#1}\nlm_{#1}}{\e\nlm}
}
\NewDocumentCommand{\en}{e_}{
\IfValueTF{#1}{\e_{#1}n_{#1}}{\e n}
}
\NewDocumentCommand{\nlk}{e_}{
\IfValueTF{#1}{n_{#1}l_{#1}k_{#1}}{nlk}
}
\NewDocumentCommand{\enlk}{e_}{
\IfValueTF{#1}{\e_{#1}\nlk_{#1}}{\e\nlk}
}
\NewDocumentCommand{\enl}{e_}{
\IfValueTF{#1}{\en_{#1}l_#1}{\en l}
}
\NewDocumentCommand{\nl}{e_}{
\IfValueTF{#1}{n_{#1}l_#1}{n l}
}

\NewDocumentCommand{\nnl}{s}{
\IfBooleanTF{#1}{n_1 n_2 l}{n_1; n_2; l}
}
\NewDocumentCommand{\ennl}{s}{
\IfBooleanTF{#1}{\en_1 \en_2 l}{\en_1; \en_2; l}
}

\NewDocumentCommand{\gslm}{s}{
\IfBooleanTF{#1}{\sigma\lambda\mu}{\sigma;\lambda\mu}
}

\newcommand{\bk}{\mathbf{k}}

\newcommand{\D}[1]{\operatorname{d}{#1}} 
\newcommand{\gs}{\ensuremath{g_\text{s}}}
\newcommand{\gb}{\ensuremath{g_\text{b}}}
\newcommand{\nmax}{\ensuremath{n_\text{max}}}
\newcommand{\lmax}{\ensuremath{l_\text{max}}}

\newcommand{\DOS}{\mathrm{DOS}}
\newcommand{\LDOS}{\mathrm{LDOS}}

\newcommand{\Eband}{\epsilon_{\text{band}}}
\newcommand{\Efermi}{\epsilon_{F}}
\usepackage{graphicx}%
\usepackage{dcolumn}%
\usepackage{bm}%

\usepackage[names]{xcolor}
\newcommand{\revision}[1]{{#1}}

\begin{document}

\preprint{APS/123-QED}

\title{Learning the electronic density of states in condensed matter}%
\author{Chiheb Ben Mahmoud$^*$}%
\thanks{These authors contributed equally to this work}

\author{Andrea Anelli$^*$}
\affiliation{%
Laboratory of Computational Science and Modeling, IMX, \'Ecole Polytechnique F\'ed\'erale de Lausanne, 1015 Lausanne, Switzerland
}%

\author{G\'abor Cs\'anyi}
\affiliation{
 Engineering Laboratory, University of Cambridge, Trumpington Street, Cambridge CB21PZ, United Kingdom%
}%

\author{Michele Ceriotti}
\email{michele.ceriotti@epfl.ch}
\affiliation{%
 Laboratory of Computational Science and Modeling, IMX, \'Ecole Polytechnique F\'ed\'erale de Lausanne, 1015 Lausanne, Switzerland
}%

\date{\today}%

\begin{abstract}
The electronic density of states (DOS) quantifies the distribution of the energy levels that can be occupied by electrons in a quasiparticle picture, and is central to modern electronic structure theory. It also underpins the computation and interpretation of experimentally observable material properties such as optical absorption and electrical conductivity.
We discuss the challenges inherent in the construction of a machine-learning (ML) framework aimed at predicting the DOS as a combination of {\em local contributions} that depend in turn on the  geometric configuration of neighbours around each atom, using quasiparticle energy levels from density functional theory as training data.
We present a challenging case study that includes configurations of silicon spanning a broad set of thermodynamic conditions, ranging from bulk structures to clusters, and from semiconducting to metallic behavior. We compare different approaches to represent the DOS, and the accuracy of predicting quantities such as the Fermi level, the electron density at the Fermi level, or the band energy, either directly or as a side-product of the evaluation of the DOS. We find that the performance of the model depends crucially on the resolution chosen to smooth the DOS, and that there is a tradeoff to be made between the systematic error associated with the smoothing and the error in the ML model for a specific structure. We find however that the errors are not strongly correlated among similar structures, and so the average DOS over an ensemble of configurations is in very good agreement with the reference electronic structure calculations, despite the large nominal error on individual configurations. 
We demonstrate the usefulness of this approach by computing the density of states of a large amorphous silicon sample, for which it would be prohibitively expensive to compute the DOS by direct electronic structure calculations, and show how the atom-centered decomposition of the DOS that is obtained through our model can be used to extract physical insights into the connections between structural and electronic features.

\end{abstract}

\pacs{Valid PACS appear here}%
\maketitle

\section{\label{sec:intro}Introduction}

The combination of electronic structure calculations and machine learning (ML) techniques has become commonplace in the atomistic modelling of matter. In particular, this combination has proven very valuable for building models that can inexpensively predict, using only atomic coordinates as inputs, any property that can be computed by first-principles calculations, based on a small number of reference calculations~\cite{Behler2016, butl+18nature}. The earliest and now most widespread applications have focused on the construction of interatomic potentials by predicting total energies and atomic forces
~\cite{behler2007, bartok2010, Li2015}, reducing dramatically the cost of \textit{ab initio} simulations. 
Since the early demonstrators, general and transferable ML potentials have been built to describe the potential energy surface of a material across many phases and including a wide variety of defects~\cite{Bartk2018}, and have also become prominent in studying amorphous materials~\cite{DeringerAdvMat}.
These efforts go beyond the more traditional fitting of semi-empirical potentials, as they aim to describe in a completely general, possibly non linear, manner the correlations between atomic configurations and target properties. Recently, machine-learning models have also been used to learn and predict other properties of crystals and molecules such as ionisation energies~\cite{Rupp2015}, NMR chemical shifts
~\cite{Cuny2016, Paruzzo2018}, dielectric response properties~\cite{wilk+19pnas}, as well as properties that are more directly linked to  electronic structure, such as the charge density~\cite{broc+17nc, gris+19acscs, alre+18cst, Chandrasekaran2019, fowl+19jpm} or the position of the Wannier centers~\cite{zhang-wannier_centers}.
Building models of electronic properties can be an end in itself, or be useful as a stepping stone to evaluate quantities that have a clear physical relation to the predicted quantity, e.g. for the electron density the exchange-correlation energy~\cite{broc+17nc}, the electrostatic potential~\cite{fabr+19cs}, or topological descriptors of chemical bonding~\cite{gris+19acscs}.

The electronic density of states (DOS) is another quantity that underlies many useful materials properties. It can be used to calculate the electronic contribution to the heat capacity in metals, the density of free charge carriers in semiconductors, and is an indirect proxy for properties such as the energy band gap, the band energy and also the optical absorption spectrum. 
In recent years, several approaches have been proposed to ``learn'' the DOS of materials, with applications including the prediction of the DOS value at the Fermi energy in alloys ~\cite{Schtt2014}, the main features of the DOS curves of transition metals~\cite{Broderick2011} or the local density of states of structures from ab-initio molecular dynamics trajectories~\cite{chan+19npjcm}. However, these attempts have been restricted to predicting particular values of the DOS curve, or rather limited in the diversity of structures included in the training and testing.

In this work, we present a machine-learning framework aimed at predicting the DOS, based on Gaussian process regression (GPR) and on a representation of the local geometry using a description of the atomic environments, known as the Smooth Overlap of Atomic Positions (SOAP)~\cite{bartok2013, de+16pccp}, the same that is widely used to build interatomic potentials and is just one example of a set of closely related representations based on the atomic neighbour density
~\cite{ceri+18book,will+18jcp,drau19prb}.
We investigate in depth the impact of different approaches to represent the DOS as a target of the ML model, and highlight the strong impact of the level of Gaussian smearing applied to the DOS on the difficulty of the regression task. We also assess the relative performance of models that directly predict  electronic properties such as the band energy and the optical absorption spectrum, with those of models that use the DOS as an intermediate quantity. All these results are presented together with uncertainty quantification based on a committee model~\cite{musi+19jctc} that provides an assessment of the quality of the predicted DOS and its derived quantities. 
We use as benchmark a challenging data set of silicon structures that includes different solid phases, liquid and amorphous configurations, and gas phase clusters, spanning a wide range of behaviors from metallic to semiconducting. Finally, we demonstrate the transferability of the model to predict the DOS of very large amorphous configurations, exploiting the local description to link atomic environments to their contributions to the overall density of energy levels. 

\section{\label{sec:methods}Methods}

\begin{figure*}
    \centering
    \includegraphics[width=1.\textwidth]{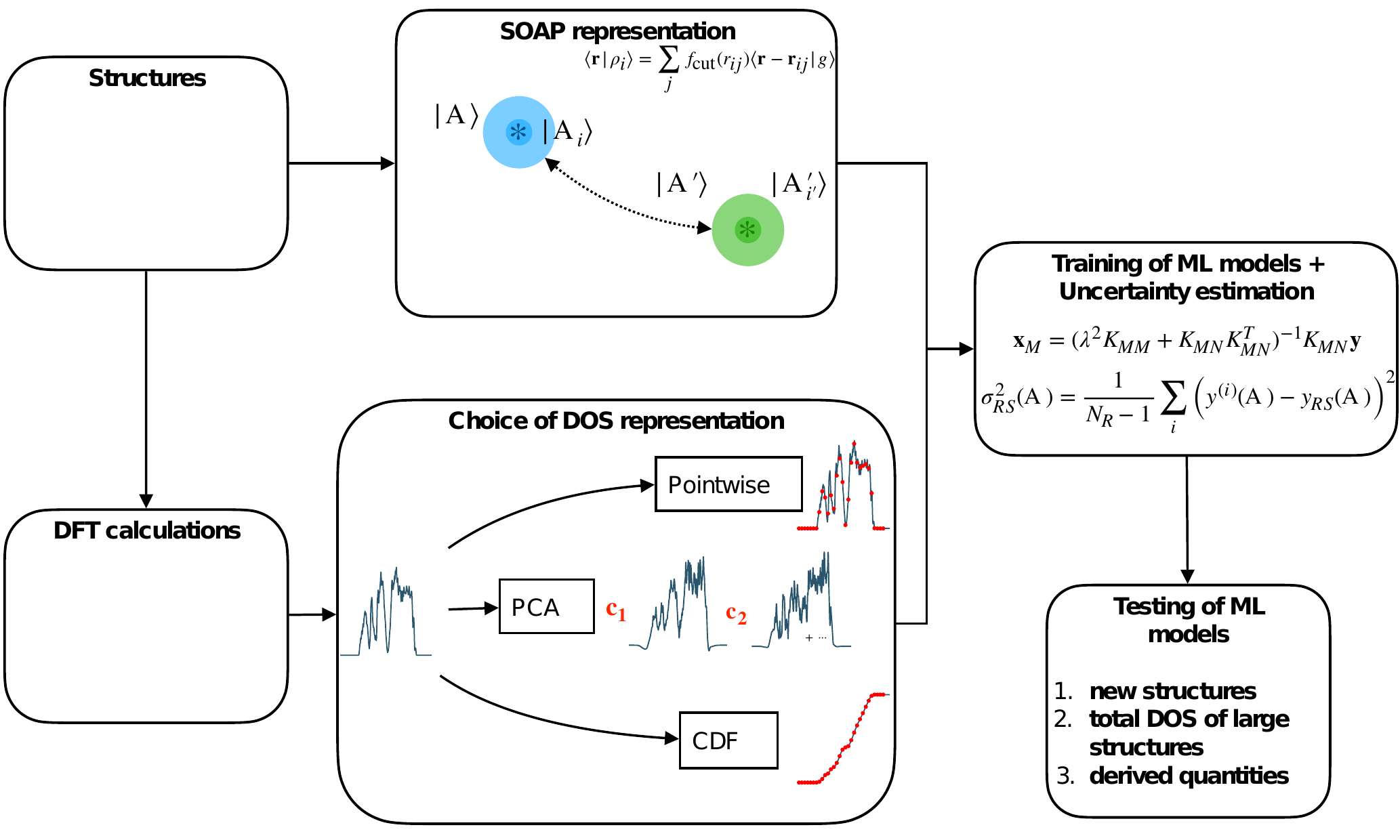}
    \caption{
    Schematic representation of the DOS ML workflow. $A$ indicates full atomic configurations, $A_i$ represents the atomic environments in these structures, the blue curves are the DOS from DFT calculations and the red dots are the targets of the ML models.} 
    \label{fig:workflow}
\end{figure*}

We define the DOS as a sum of Dirac distributions centered around the eigenvalues of the single-particle Hamiltonian, $E_n(\bk{})$ describing the energy levels accessible to  electrons can occupy at each point $\bk$ of the electronic Brillouin zone (BZ):
\begin{equation*}
    \DOS(E) = \frac{2}{N_k} \sum_n^{bands} \sum_{\bk{}} \delta(E-E_n(\bk{})).
\end{equation*}
This definition approximates the integral over the BZ, and extends readily to a-periodic systems by removing the summation over $k$ points.
In order to obtain a continuous distribution, the Dirac distributions are broadened with a Gaussian smearing $\gb$: $\delta(E-E_n(\bk{})) \rightarrow (\gb\sqrt{2\pi})^{-1}\exp[-(E-E_n(\bk{}))^2/(2\gb^2)]$.

In the following subsections we introduce the different components of our strategy to predict the DOS based on a regression model trained on a small number of reference configurations. 
We discuss briefly the equations that underlie a Gaussian process regression model that includes an inexpensive estimate of the prediction uncertainty, different possible choices to discretize the density of states, and the family of representations that we use to describe atomic environments (Fig. (\ref{fig:workflow})). 

\subsection{\label{app:mlmodel}An atom-centered model of the DOS}

We aim to build a model of the DOS for a structure $A$
by decomposing the total DOS into a sum of local contributions from each of its atomic environments $A_i$, i.e. 
\begin{equation}
\DOS(A,E) = \sum_{i\in A} \LDOS(A_i,E).
\label{eq:dos-additive}
\end{equation}
Implementing such a model requires the definition of a framework to parameterise the shape of the LDOS, and a framework to represent the structure and the composition of the  environment surrounding each atom. Given these, one can determine the parameters $\bx$ of the LDOS model by minimising a loss function of the form:
\begin{equation}
\ell^2(A, \bx) = \int \mathrm{d}E 
\left| 
\DOS(A,E) - \sum_{i\in A} \LDOS_{\bx}(A_i,E) 
\right|^2.
\label{eq:loss-generic}
\end{equation}
The model can then be used to make predictions for new, possibly more complex, structures. 
In this work we model $\LDOS_{\bx}(A_i,E)$ using the Projected Process (PP) approximation of the Gaussian Process Regression (GPR)~\cite{rasm06book}. We use the most diverse $M$ atomic environments, selected according to a Farthest Point Sampling (FPS) scheme among the training structures, as the \textit{active set} that defines the basis on which the target is expanded in local contributions:
\begin{equation}
     \LDOS_\bx(E, A_i) = \sum_{j\in M} x_j(E) k(A_i,M_j ),
\end{equation}
where $k(A_i, M_j)$ is a positive-definite kernel basis function that expresses the similarity between the environment $A_i$ and an environment from the active set $M_j$, that will be defined in Section~\ref{app:soap}. 
Given the additive nature of the DOS model~\eqref{eq:dos-additive}, we define the kernel between entire structures as the sum of the kernels between the atomic environments that constitute the structures, $k(A,A')=\sum_{i\in A,i'\in A'} k(A_i,A'_{i'})$.
The linear expansion coefficients $x_j(E)$ depend on the energy and should be discretised in a way that reflects the representation of the DOS, which is discussed in Section~\ref{app:dosrepr}. 
We optimise the coefficients $\bx_M(E)$ by minimising the following empirical loss function, in the same spirit of ridge regression models, based on knowledge of the targeted DOS for the training structures:
\begin{equation}
\ell^2_\lambda(\bx_M) = \sum_{A \in \text{train}}
\ell^2(A,\bx_M) + \lambda^2 \bx^T_M K_{MM} \bx_M.
\end{equation}
Here, $\lambda$ is the regularisation parameter and $K_{MM}$ is the kernel matrix, whose entries are the kernel functions between the active set environments. The optimal solution to this problem is obtained as a function of the kernel matrix of the active set $K_{MM}$ and the kernel matrix of the training structures and the active set $K_{NM}$:
\begin{equation}
    \bx_M(E) = (\lambda^2 K_{MM} + K_{MN} K_{MN}^T)^{-1} K_{MN} \mathbf{y}_N(E),
\end{equation}
where $\mathbf{y}_N$ is a vector containing the values of $\DOS(A, E)$ for the $N$ training structures. Once we find the optimal solution for our problem, usually using a k-fold cross-validation scheme, the DOS of a new structure $A_*$ can be obtained as a simple dot product:
\begin{equation}
    \DOS(A_*, E) = \bk{}_{A_*M}^T \cdot \bx_M(E),
\label{eq:pp-prediction}
\end{equation}
where $\bk{}_{A_*M}$ is the vector that contains the kernels between the structure $A_*$ and the $M$ active set environments.

\subsection{\label{app:uncertainty}Uncertainty estimation}

Gaussian process regression models have a built-in variance estimator, that makes it possible to assess the statistical uncertainty -- and hence the reliability -- of the prediction for a specific structure~\cite{rasm05book}. For computational efficiency, and to simplify the propagation of uncertainty from the atom-centered decomposition to the full density of states of a structure, we build instead a committee of $N_{RS}$ GPR models of size $n<N$, as discussed e.g. in Ref.~\citenum{musi+19jctc}. If the models are built based on the PP approximation, keeping a fixed active set for all the models, each model corresponds to a different weight vector $\bx_M^{(i)}(E)$, and predictions can be obtained inexpensively as the kernel vector in Eq.~\eqref{eq:pp-prediction} must be computed only once for each new structure. 
The average of the predictions  $\DOS^{(i)}(A_*, E)$  made by the models in the committee is taken as the best estimate:
\begin{equation*}
\DOS_{RS}(A_*, E) = \frac{1}{N_R} \sum_i \DOS^{(i)}(A_*, E),
\end{equation*}
while their variance is taken as a measure of the uncertainty
\begin{equation}
\sigma_{RS}^2(A_*) = \frac{\alpha_{RS}}{N_R-1} \sum_i \left(\DOS^{(i)}(A_*, E)- \DOS_{RS}(A_*, E)\right)^2.
\end{equation}
The factor $\alpha_{RS}$ serves to compensate for the correlations that are present between the training points, and between the different resampled models; $\alpha_{RS}$ can be determined by calibrating the uncertainty estimate with a likelihood maximization criterion, using a validation set or an internal reference.\cite{musi+19jctc}
Note that it is possible to realize this calibration process by rescaling the predictions around the mean, i.e. 
\begin{equation}
\begin{split}
\DOS^{(i)}(A_*, E) \leftarrow &\DOS_{RS}(A_*, E) + \\ & \sqrt{\alpha_{RS}} \left[ \DOS^{(i)}(A_*, E) - \DOS_{RS}(A_*, E) \right].
\end{split}
\end{equation}
We can use this calibrated ensemble of predictions to perform post-processing tasks, such as the assembly of atom-centered predictions, in a way that automatically incorporates correlations between the predictions of different models. 

\subsection{\label{app:dosrepr}DOS representation}

In order build a practical model of the DOS, one needs to represent the (L)DOS as a set of discrete values, $y_j(A)$, that can then be used to construct a multivariate regression model. A pointwise, trivial approach is to discretise the energy axis over a finite range $\left[E_0,E_0+(N_{\text{DOS}}-1)\delta E\right]$, and take the smooth DOS computed at each energy point as a regression target,
\begin{equation}\label{eq:dos-trivial}
y_j^{\mathrm{PW}}(A) = \DOS(E_0 + j\delta E, A).
\end{equation}
In the Gaussian process regression model we use, each regression task is independent of the others, which means that the number of these descriptors can become arbitrarily high depending on the level of complexity (defined by the Gaussian broadening) and the density of the considered energy points (linked to the discretization width $\delta E$), which results in an increase in the number of prediction models.

This pointwise representation is not necessarily the most efficient: it potentially requires to train and evaluate hundreds of ML models, and it ignores the fact that variations in the DOS between different structures and different energy levels are correlated, both because of physical reasons and because of the Gaussian broadening. 
It is possible to reduce the degrees of freedom of the problem  by projecting the DOS on an orthogonal basis set, and learning the expansion coefficients. 
In order to capture the correlations between the variations of the DOS at different points, we construct a data-adapted basis by evaluating the principal components (PC) of the DOS within the training set.
Given that we aim to build a predictor for the LDOS, but we only have information on the total DOS of a structure, we normalize the DOS of each structure by the number of electronic states, and construct the matrix
\begin{equation}
\tilde{Y}_{A j} =  
\frac{y_j(A)}{N_{A}} - \frac{1}{N_\text{train}} \sum_{A^\prime} \frac{y_j(A^\prime)}{N_{A^\prime}}.
\end{equation}
We then compute the eigendecomposition of the covariance matrix
\begin{equation}
\mathbf{Y}^T\mathbf{Y}= 
\mathbf{U} {\boldsymbol{\Lambda}}  \mathbf{U} ^T.
\end{equation}
The columns of the unitary matrix $\mathbf{U}$ that are associated to the largest eigenvalues $\Lambda_k$ describe uncorrelated modes of variation of the (L)DOS. 
The truncation of the expansion to a small number of  PCs  determines the error that one makes in approximating the DOS, and corresponds effectively to an additional smoothing of the DOS. 
Building a model in the PC representation amounts to computing the projection of the DOS on the basis functions
\begin{equation}\label{eq:dos-pc}
\tilde{y}^\text{PC}_k(A) = \sum_j y_j(A)\ U_{jk},
\end{equation}
training a regression model on each of the $\tilde{y}^\text{PC}_k$ coefficients and then reconstructing the prediction in terms of the principal vectors,
\begin{equation}
y_j(A) \approx \sum_k \tilde{y}^\text{PC}_k(A)\ U_{jk},
\end{equation}

A third approach to represent the DOS can be derived to address the fact that a loss of the form \eqref{eq:loss-generic} cannot discriminate between distributions that differ by the position of peaks that have negligible overlap -- a problem that is frequently encountered when comparing spectral functions. 
The Wasserstein distance (also known as earth mover's distance) is a metric to compare distributions designed to address this problem, and that can be easily computed as the norm of the pointwise difference between the inverse cumulative distribution functions~\cite{Rubner}. 
Inspired by the Wasserstein metric, we propose to represent the DOS in terms of the associated cumulative distribution function (CDF), that can be computed as partial sums over the pointwise representation, which approximate the integral over the energy:
\begin{equation}\label{eq:dos-cdf}
y^\text{CDF}_k(A) = \sum_{j=0}^k y_j(A).
\end{equation}
Even though a Euclidean norm that uses this vector is \emph{not} a precise implementation of the earth mover's distance (that is based on the Euclidean distance between \emph{inverse} CDFs), it is sensitive to shifts in peak position, and preserves the additive construction of the total DOS based on atom-centered contributions -- a physical constraint that would be lost by using a metric based on the inverse CDF. 

\subsection{\label{app:soap}Structural representation}

We describe atom-centered environments using a representation that corresponds to a smoothed version of a 3-body correlation function~\cite{will+19jcp}, that we indicate as the vector $\rep|A; \frho_i^2>$. \revision{We use the notation introduced in Ref.~\citenum{will+19jcp}, in which feature vectors for an atom-centered environment $A_i$ are indicated with $\rep<X||A_i;\text{rep.}>$ -- where $X$ indicates a set of indices that enumerate the components, and ``rep.'' is a shorthand that describes the kind of correlation that underlies the featurization (e.g. $\frho_i^2$ indicates the two-point density correlation).}
In practice, we use the projection of this correlation function on a basis of radial functions $R_n(x)\equiv \rep<x||n>$ and spherical harmonics $Y_l^m(\bxhat)\equiv \rep<\bxhat||lm>$, that corresponds to the SOAP power spectrum introduced in Ref. ~\citenum{bart+13prb}.
The SOAP feature vector can be computed by first expanding  the Gaussian smeared atomic density centered on the $j^{\mathrm{th}}$ atom (restricted by a cutoff function to within a radial cutoff distance $r_c$) on an orthogonal basis,
\begin{equation}
\begin{split}
\rep<\br||A;\frho_i> =  &
\sum_{j\in A} f_{\mathrm{cut}}(r_{ji}) \rep<\br-\br_{ji}||g>,\\ %
\rep<nlm||A; \frho_i> = &\int \D{\bx} 
\rep<n||x> \rep<lm||\bxhat> \rep<\bx||A;\frho_i>
\end{split}
\end{equation}

We indicate with $\gs$ the standard deviation of the Gaussian $g(\bx)\equiv\rep<\bx||g>\equiv \exp(-\bx^2/2\gs^2)$, with $\nmax$ the number of radial basis functions, and with $\lmax$ the maximum angular momentum channel.
Based on these expansion coefficients, one can build a hierarchy of rotationally-invariant features. The second-order invariant corresponds to the SOAP power spectrum,
\begin{equation}
\rep<\nnl*||\frho_i^2> = \frac{1}{\sqrt{2l+1} } \sum_m (-1)^m \rep<n_1lm||\frho_i> \rep<n_2l\,(-m)||\frho_i>,
\end{equation}
where we omit the indication of the structure $A$, for simplicity.
The scalar product between the power spectrum vectors can be used to define the SOAP kernel we use in this work:
\begin{equation}
    k(A_i,A'_{i'}) \propto  \rep<A; \frho_i^2||A'; \frho_{{i'}}^2>^\zeta.
\end{equation}
An exponent $\zeta>1$ effectively introduces non-linear terms that corresponds to some (but not all~\cite{pozd+20prl}) higher body order correlations. 
In order to better reflect the locality of the LDOS, and the contribution of neighbors to the structure-property relation, 
we use a radial scaling of the SOAP features that is implemented as an additional weighting of the contributions from the neighbors, $u(r_{ij}) = \frac{c}{c+(r_{ij}/r_0)^m}$, where $c$, $m$ and $r_0$ are parameters to be optimized with respect to the target property of the learning scheme. An optimized radial scaling can improve substantially the performance of a model, similarly to what can be obtained by the use of multiple kernels with different length scales~\cite{will+18pccp}. %

\section{\label{sec:computational}Computational details}

\begin{figure}
    \centering
    \includegraphics[width=.45\textwidth]{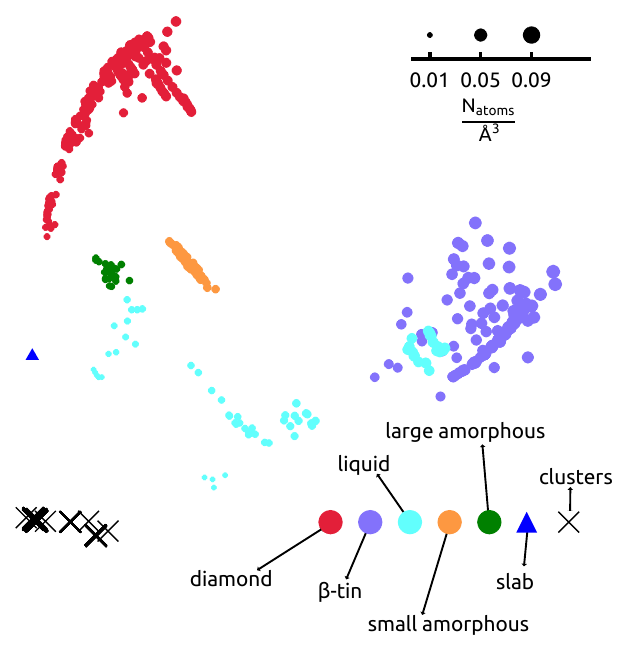}
    \caption{Clustering of the structures in the Silicon data set based on the first 2 kernel principal components of  every configuration. These components hold $\approx94\%$ of the variance in features space. The different subsets are well separated in  features space, except for a few liquid structures computed at high pressure, that partly overlap with $\beta$-tin phase configurations. \revision{We also plot, on the same axes, the position of carefully-equilibrated, large a-Si supercells (discussed in Section~\ref{sec:a-si}) and of one Si(100) slab (discussed in Fig.~\ref{fig:dos_slab}), used to assess the extrapolative capabilities of the model.} }
    \label{fig:kpca}
\end{figure}

We use as a training and validation data set a challenging combination of 1039 silicon structures containing configurations that correspond to elastically and thermally distorted bulk diamond and $\beta$-tin structures, snapshots from molten silicon simulations, amorphous configurations obtained at different quenching rates, as well as some cluster configurations. We extract these structures from the training data set used to build a machine-learning interatomic-potential for silicon\cite{Bartk2018}. 
Fig.~\ref{fig:kpca} demonstrates the heterogeneity of the data set, showing a projection on the two largest principal components of the average SOAP power spectrum vectors of the different configurations. The parameters of the SOAP are the same as those used for the regression models, discussed below. The map reflects the presence of several distinct groups of structures, that have been obtained with simulations performed at different temperatures and pressures. 
In what follows we randomly selected  800 structures that we used to train the different models, and used the remainder of the data set for testing.  \revision{The random selection ensures approximately uniform sampling of the different portions of the configuration space. Unless otherwise specified, test errors are averaged over 16 random splittings of the overall data set.}

We compute the single-particle energy levels for this system by running density-functional theory calculations using the FHI-aims all-electrons code~\cite{Blum2009}.
We use the ``tight'' settings, and the PBE~\cite{Perdew1996} exchange-correlation functional. We keep a constant k-point spacing of 0.01\AA$^{-1}$ for the periodic structures.
\revision{The energy levels are aligned by zeroing the vacuum level of the Hartree potential for isolated structures, and its constant, $G=0$ component for periodic structures.} 
As discussed below, we compute the density of states by summing over the single-particle eigenvalues, using a Gaussian smearing with various widths.

As introduced in Section \ref{app:soap}, the SOAP representation has several hyperparameters that need to be tuned depending on the training data and the target property.
Given that, as discussed above, we aim to compare the performance of the model with different representation of the DOS and different target properties, one would need to perform hundreds of separate optimization procedures for these hyperparameters. 
Instead, we performed a single optimization using the cohesive energy as the target property; this avoids biasing explicitly our comparison towards one of the  DOS learning protocols, and is representative of a scenario in which one wants to re-use the features that underlie a machine-learning potential to estimate additional electronic-structure properties. 

We use the metric induced by a preliminary set of SOAP features to select, by farthest point sampling (FPS)~\cite{imba+18jcp}, 3000 environments \revision{ out of $\approx$22000 environments} to use as active points for the PP approximation. As shown in the SI, increasing the active set size further leads to negligible reduction of the prediction error.
We then selected the best hyperparameters using a 5-fold cross validation regression scheme and a grid search, using a single random ordering of the full data set. We obtain the smallest prediction errors for the cohesive energy for the following set of parameters: $r_c=6\text{\AA}$, $\nmax=12,\ \lmax=9,\ \gs=0.45,\ c=1,\ m=5$, and $r_0=3.0\text{\AA}$.
As shown in the SI, this choice of parameters lead only to a marginal degradation of performance in comparison to a model that has been specifically optimised to reproduce the DOS. 
For consistency, given that the change of hyperparameters modifies the kernel, and the kernel-induced distance, we then re-selected 3000 active environments using these optimal values. It should be noted, however, that selecting new active points led to negligible improvement of model accuracy.

We built 8 models using the same active set but different 50\%{} subsampling of the 800 train structures. 
We report the mean of the model as our best predictions (which has an accuracy comparable to that of a single model trained on the full 800 structures), and rescale the spread of the models around the mean, as discussed in Section~\ref{app:uncertainty}, to obtain an ensemble of predictions based on which we can easily propagate our uncertainty quantification.

\begin{figure}
    \centering
    \includegraphics[width=1.0\linewidth]{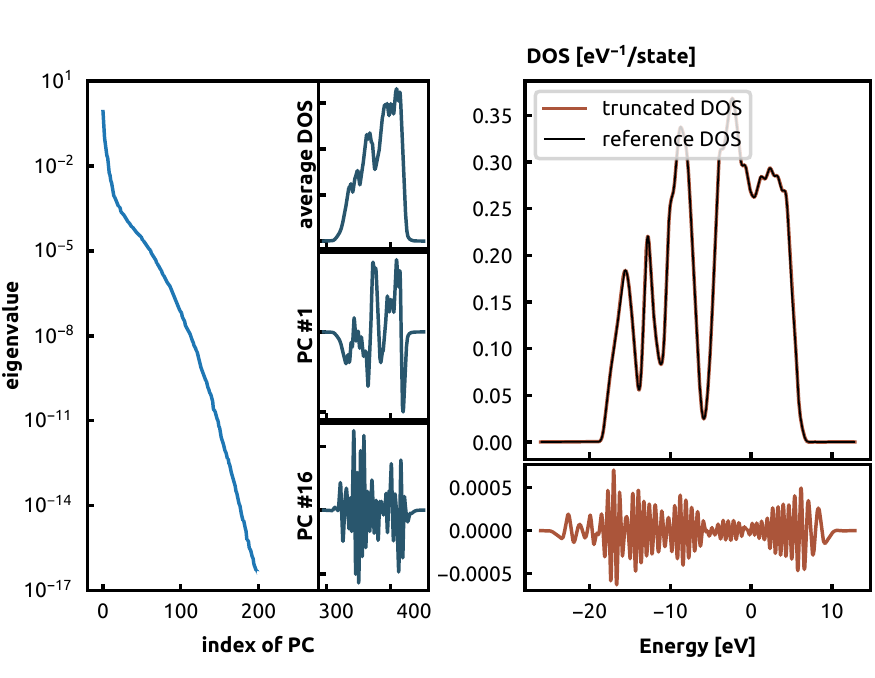}
    \caption{(Left) Distribution of the first 200 eigenvalues of the covariance matrix of the DOS at a $\gb=0.3$~eV. The 3 small panels on the right represent the shape of the average DOS in the data set, the $1^{st}$ principal component and the $16^{th}$ principal component. (Right) The reference DOS curve of a Silicon diamond structure at $0.3$~eV smearing and its reconstruction from the first 80 PCs. The lower panel shows the errors at every energy level. The total error for this structure is $\approx 5.11\times 10^{-3}$ $\mathrm{eV^{-1}}/$atom.}
    \label{fig:pca_decompo}
\end{figure}

In order to investigate the impact of the representation of the DOS on the performance of the model, we consider three values of the Gaussian broadening $\gb$: $0.1$~eV, $0.3$~eV and $0.5$~eV. We discretise the DOS on a grid where the points are spaced by $\delta E=0.05$~eV, which ensures that we are able to sample the fine structure of $\DOS(E)$ when using a $\gb=0.1$eV smearing. We use this representation as the trivial representation, Eq.~\eqref{eq:dos-trivial}. 
We use the same grid to compute the cumulative integral of the DOS, Eq.~\eqref{eq:dos-cdf}. 
For the PC representation, Eq.~\eqref{eq:dos-pc}, we select the principal eigenvectors of the covariance matrix computed for the 800 training structures. 
The left panel in Fig.~\ref{fig:pca_decompo} demonstrates the rapid decay of the eigenvalues of the covariance for  $\gb=0.3$~eV, and the shape of the average DOS of the data set, the $1^{st}$ and the $16^{th}$ eigenvectors. One notices that the the principal components corresponding to the high eigenvalues contribute to the general structure of a DOS curve, while the ones corresponding to lower eigenvalues describe the fine structure.
We choose the number of PCs to retain 99.99\%{} of the variance, which corresponds to 200, 80 and 35 PCs for $\gb=0.1, 0.3, 0.5$~eV, respectively.
Even though the error resulting from this approximation is visually very small, as shown in the right panel of Fig.~\ref{fig:pca_decompo}, it leads to non-negligible errors when using the DOS to compute derived properties such as the band energy.

\section{\label{sec:results}Results}

\begin{figure*}
    \centering
    \includegraphics[width=0.95\linewidth]{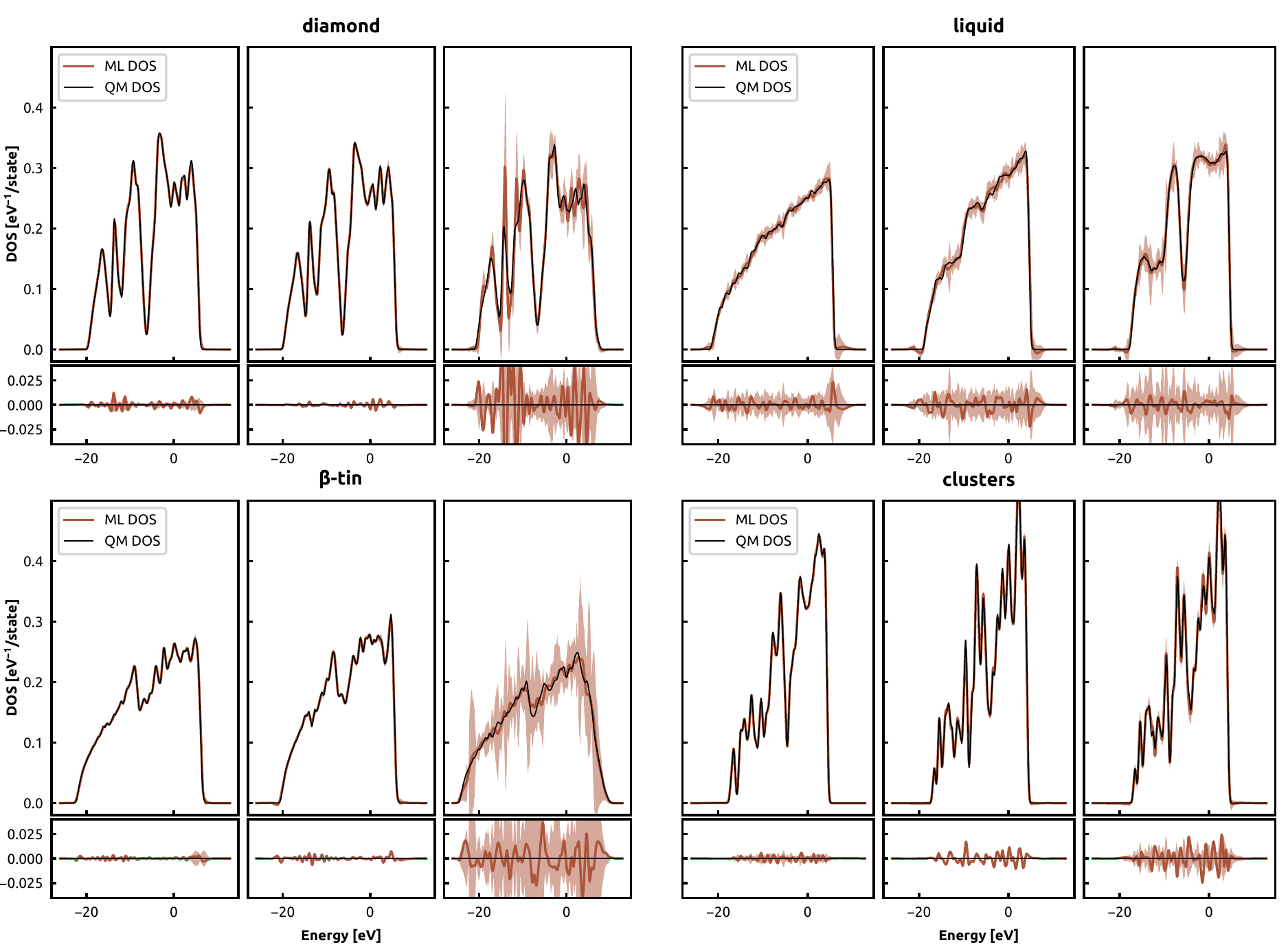}
    \caption{Representative examples of DOS predictions for the Silicon data set. The 4 panels represent one of the subsets in the data set: diamond, liquid, $\beta$-tin structures and clusters. Every panel shows three cases corresponding to the best, median and worst predicted uncertainty in the test set, compared to the reference DFT DOS  (left to right). The reference DOS is constructed using $\gb=0.3$eV. The shaded areas indicate the uncertainty estimates of the ML model at every energy level, based on a committee of 8 GPR models. The lower section of each plot depicts the residuals, color-coded in the same way as the plot of the prediction. }
    \label{fig:dos_ex}
\end{figure*}

Having discussed the details of the DOS model, we now turn to assessing its performance on the database of silicon structures. We begin by comparing the accuracy as a function of the smearing of the density of states, and its representation, and proceed to determine how the error in the prediction of $\DOS(E)$ translate to the error in quantities that can be obtained from it, such as the band energy or the Fermi level. 

To facilitate the comparison of the model performance between different properties and scenarios we normalise the root mean squares error (RMSE) by the standard deviation of the target property, expressed as a percentage. This is particularly important because reducing the Gaussian smearing increases substantially the the complexity of the DOS, measured in terms of the variance over the full data set (see Table S1 in the SI). For scalar properties, the expressions reads:
\begin{equation}
\mathrm{RMSE}_{\mathrm{scalar}} = 100 *\frac{ \sqrt{\frac{1}{N} \sum (y_{pred}^i-y^i)^2}}{ \sqrt{\frac{1}{N} \sum (y^i - \bar{y})^2}}
\end{equation}
This expression is easily extended to cover the properties that have an energy dependence \revision{(such as the DOS and the distribution of excitations)}, that require the simultaneous regression of multiple coefficients, by comparing the $L^2$ distance to the deviation from the average vector representing the target property of the training set:
\begin{equation}\label{eq:vect_errors}
\mathrm{RMSE}_{\mathrm{vec}} = 100 * 
\frac{\sqrt{\frac{1}{N} \sum_i \int (\mathbf{y}_{pred}^i-\mathbf{y}^i)^2}}
{\sqrt{\frac{1}{N} \sum_i \int (\mathbf{y}^i - \bar{\mathbf{y}})^2}},
\end{equation}
where $\mathbf{\bar{y}}$ is the vector containing the average of each coefficient in the $\mathbf{y}^i$. We will use either of the definitions as appropriate, and indicate the error simply as \%RMSE.

\subsection{\label{res:model_performance}\revision{Comparison of DOS representations}}

\begin{figure}[tbp]
	\centering
	\includegraphics[width=1.0\linewidth]{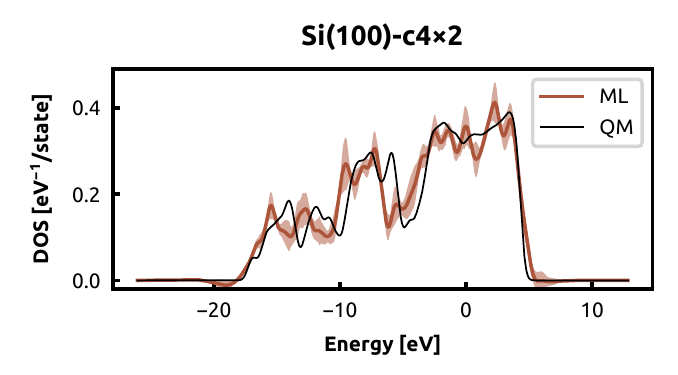}
	\caption{\revision{Comparison of DFT and ML DOS of 96-atom slab model of the  Si$(100)-c4\times 2$ surface reconstruction. The reference DOS is constructed using $\gb=0.3$eV. The ML model use the pointwise representation of the DOS. The shaded area represents the uncertainty of the ML model at each energy level.}}
	\label{fig:dos_slab}
\end{figure}

We begin by showing, for the pointwise representation of the DOS and $\gb=0.3$eV, a plot of the model ${\rm DOS}(E)$ for the diamond and liquid structures with the lowest, median and highest predicted uncertainty (Figure~\ref{fig:dos_ex}).
The figure demonstrates that a  single model is capable of predicting the behavior of Si across the semiconductor to metal transition, and that the uncertainty quantification correctly identifies the most problematic test structures.
\revision{As an example of the stability of the model when performing extrapolative predictions, we estimate the DOS of a 96-atom Si(100) slab, with one surface truncated to the bulk geometry, and the other reconstructed with a $c4\times 2$ geometry (Fig.~\ref{fig:dos_slab}, geometry from Ref.~\citenum{ceri+09prb}), using the pointwise representation and a target Gaussian broadening of $\gb=0.3$eV. As shown in Fig.~\ref{fig:kpca}, this structure is isolated in phase space, which results in both the predicted uncertainty and the actual error being large, comparable to the worse-case scenarios in Fig.~\ref{fig:dos_ex}.
Even in this challenging example, however, the qualitative features of the DOS are correctly reproduced, and the large uncertainty could be used in an active learning setting to signal the need for a refinement of the training set. }

\begin{figure}[tbp]
    \centering
    \includegraphics[width=0.8\linewidth]{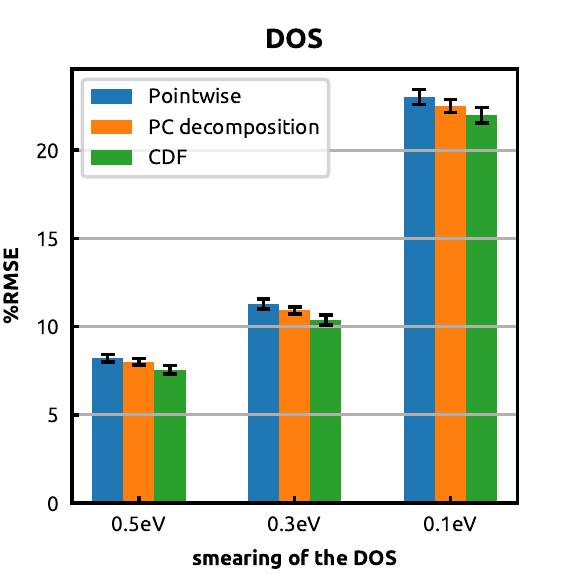}
    \caption{ \revision{ Average errors in the DOS over 16 train/test splits in the Si data set using 3 different representations of the DOS curves: the pointwise approach, the decomposition on a basis set of principal components, and the DOS derived from the learnt CDF, and for 3 different Gaussian broadening values: $0.5$eV, $0.3$.eV and $0.1$eV. Different representations lead to comparable errors, that grow systematically with decreasing $\gb$. The error bars represent the standard error of the mean. Errors for the PC decomposition due to the truncation of the PC basis are negligible.}  }
    \label{fig:dos_errors}
\end{figure}

In order to assess more quantitatively the accuracy of the model for different values of $\gb$ and different representations of the DOS, we then compute the \%RMSE of predictions (Eq.~\eqref{eq:vect_errors}) using the full training set of 800 structures, shown in Fig.~\ref{fig:dos_errors}. 
To account for the dependency of the accuracy on the test/train split, we repeated regression and testing on 16 random splits, and report mean and standard error of the mean over the different splits. 
Even though we have renormalized the error on the intrinsic variance of the data, which is larger for the smaller values of the Gaussian smearing, the error in the predicted DOS is clearly much larger for the finer $\gb$. The errors jump from roughly $8\%$ for the $0.5$eV smearing and $11\%$ for the $0.3$eV smearing to $22\%$ for the $0.1$eV smearing.
The representation of the DOS has a small impact on the accuracy of the model, with the CDF showing a slight, but systematic, advantage over the pointwise and the PC representations. The projection errors for the latter representation are too small to affect the errors of the DOS, unlike what we will encounter later when we discuss the derived quantities.

\begin{figure}
\centering
\includegraphics{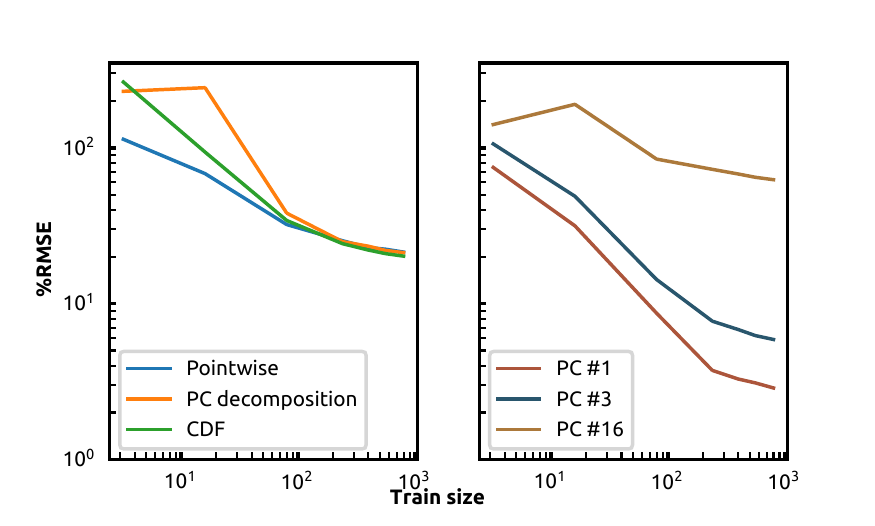}
\caption{Learning curves for (Left) the 3 different representations of the DOS curve and (Right) the projection on the $1^{st}$, $3^{rd}$ and $16^{th}$ PC as a function of the training structures. All the errors are normalised with respect to the standard deviation of the training set. The reference DOS is computed with $\gb=0.1$~eV.  }
\label{fig:lc_dos}
\end{figure}

In order to understand the limitations of our data set and to validate the training set size we use in this work, we examine the learning curves (LC) for the different representations of the DOS, as shown in the left panel of  Fig.~\ref{fig:lc_dos} for the most challenging case of $\gb = 0.1$~eV. For training sizes less than $\approx$ 100 structures, we notice that all the LCs are decreasing algebraically, and for larger sizes, the rate decreases and we can see that the LCs start to saturate, which indicates that adding more data to train the ML models does not result in a significant improvement of the models' performance.
Despite small differences at the smaller train set sizes, the three representations show similar convergence behavior. The PC  representation can provide some insight on the origin of the plateau. The right panel of Fig.~\ref{fig:lc_dos} shows the LCs of the first, third and sixteenth projections of the DOS on the PC basis, with the fractional error referring to the variance of each component.  The first two elements are well-learnt: errors are around $10\%$ for 100 structures in the training set, and the corresponding LCs saturate at low validation errors, 2\% and 8\% respectively.  In contrast with the first few principal components, the learning is slower and more difficult for lower-variance components. The error is below $60\%$ for the $16^{th}$ component only for the full training data set. In general, we observe  that the convergence of these individual errors gets slower as we consider higher PCs of the DOS. 
As shown in Fig.~\ref{fig:pca_decompo}, these smaller-variance PCs are associated with high-frequency, ``noisy'' modes, that are necessary to describe the fine structure of the DOS.
For many applications, a large Gaussian smearing does not hinder using the density of states -- and indeed previous attempts at using ML to predict the DOS used a large smearing \revision{(e.g. 0.2 eV in Ref.~\citenum{chan+19npjcm})}. Whenever a higher resolution is needed (e.g. to identify the position of individual defect states, or to determine precisely the band gap), a higher density of data, possibly in combination with more complex models, is needed. \revision{In section~\ref{sec:a-si} we show that - when focusing on a more restricted set of configurations - it is possible to achieve quantitative accuracy with a fine DOS smearing by relatively minor tuning and extension of the training set. }

\subsection{\revision{Learning from the ML DOS}}

Besides its intrinsic interest as a description of the single-particle energy levels in a condensed-phase system, the $\DOS(E)$ can be used as the starting point to derive other quantities, that relate to experimental observables.  We consider four quantities: the Fermi energy ($\Efermi$), the density of states at the Fermi energy ($\DOS(\Efermi)$), the band energy ($\Eband$), and the distribution of excitations ($A(\Delta)$), because they are easily extracted from a smooth DOS curve. For completeness, we define these quantities as follows:
\begin{itemize}
    \item The Fermi energy ($\Efermi$). 
    \begin{equation*}
    \Efermi: \int\mathrm{d}E\ \DOS(E) f_{FD}(E-\Efermi)_{|T=0} =N,
    \end{equation*}
where $f_{FD}(E)_{|T=0}$ is the occupation of the energy level according to Fermi-Dirac statistics and $N$ is the number of valence electrons.
    \item The density of states at the Fermi energy ($\DOS(\Efermi)$) 
    \item The band energy 
    \begin{equation*}
\Eband = \int \mathrm{d}E\ E\ f_{FD}(E-\Efermi)_{|T=0}\ \DOS(E).     
    \end{equation*}
    \item The distribution of excitations 
    \begin{equation*}
    \begin{split}
    A(\Delta) = &\int \int \mathrm{d}E\ \mathrm{d}E^\prime\ \DOS(E)\  f_{FD}(E-\Efermi)_{|T=0}\\& \DOS(E^\prime)\ (1-f_{FD}(E^\prime-\Efermi)_{|T=0}) \delta(E-E^\prime-\Delta).
    \end{split}
    \end{equation*}
    $A(\Delta)$ mimics the adsorption spectrum, where $\Delta$ corresponds to the absorbed photon's energy and we ignore the amplitude the transition. The shape of $A(\Delta)$ for small excitation energies reveals the presence and the magnitude of a band gap.
\end{itemize}

\begin{figure*}
	\centering
	\includegraphics[width=\textwidth]{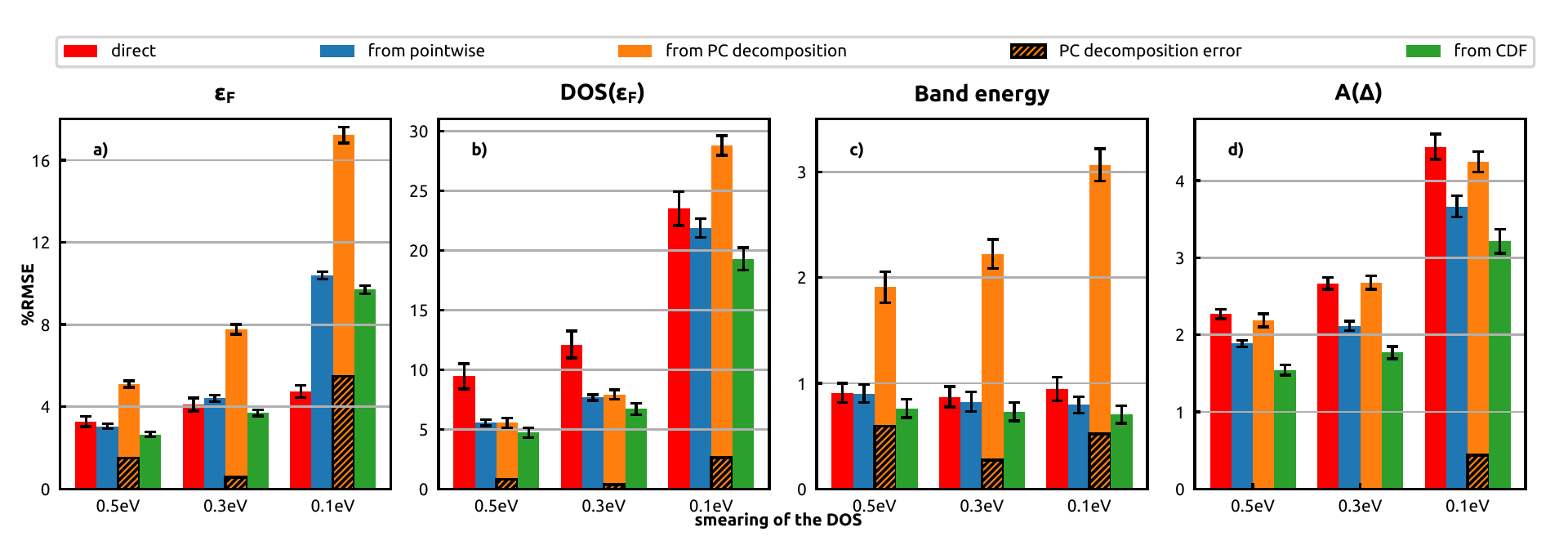}
	\caption{    Average errors in the derived quantities over 16 train/test splits in the Si data set using 3 different representations of the DOS curves: the pointwise approach, the decomposition on a basis set of principal components and derived from the learnt CDF, and for 3 different $\gb$ values: $0.5$~eV, $0.3$~eV and $0.1$~eV. The error bars represent the standard error of the mean and hatched areas represent the average systematic errors due to the projection on the basis of PCs. The test errors grow systematically when we decrease the value of $\gb$.}
	\label{fig:all_errors}
\end{figure*} 

For each of these properties we build a  ML model  using the same kernel parameters and train set, minimizing a loss function analogous to equation~\ref{eq:loss-generic}. We compare these direct predictions to \emph{indirect} models built by first predicting $\DOS(E)$ and then using it to compute $\Efermi$, $\DOS(\Efermi)$, $\Eband$, $A(\Delta)$. Whenever an expression depends on $\Efermi$, we use the value computed consistently from the predicted DOS.
We build different models with various values of $\gb$ and representation of the DOS, as in  in Section \ref{res:model_performance}. 

The average prediction errors for the four properties and for the different models, are illustrated in Fig.~\ref{fig:all_errors}.
Similar to what we observe for the DOS learning, the prediction errors increase as we decrease the value of $\gb$, including the direct method. The truncation errors in the PC decomposition representation contribute significantly to the overall errors.

Let us focus on each individual quantity, starting by the Fermi energy, whose accurate determination is particularly important, as it enters the definition of the other three quantities. The prediction errors for the direct models are low,  between $3.5\%$ and $4.5\%$. They increase, although less dramatically than in the case of the DOS, for decreasing values  of the Gaussian smearing $\gb$. Errors for the indirect predictions are comparable at  $\gb=0.3$~eV and  $\gb=0.5$~eV, except for the PC decomposition, for which the errors are almost twice as large. The truncation of the DOS contributes to the larger error as shown in the SI where a large number of PC component is necessary to obtain an accurate estimate for $\Efermi$. In combination with the difficulty in learning the fine-grained components, this explains the poor performance of the PC approach. 
Similarly to what was observed for the DOS modes, the error increases substantially for the smallest smearing value, and the direct prediction out-performs by nearly a factor of 2 all the indirect predictions.

The prediction errors of the $\DOS(\Efermi)$ of the direct model follow the same trend as the DOS, where they grow from $9.5\%$ for $\gb=0.5$~eV to $23\%$ for $\gb=0.1$~eV. In contrast to the case of the Fermi energy, here the errors of the indirect models are significantly lower for $\gb=0.5$~eV ($\approx 5\%$ error) and $\gb=0.3$~eV ($\approx 7\%$ error) and comparable between the three approaches, with a minor advantage for the CDF scheme. For $\gb=0.1$~eV errors increase substantially, but the indirect models still out-perform a direct prediction, with the exception of the PC decomposition, where the errors are close to $29\%$. The DOS truncation error contributes partly to the poor performance of the PC scheme, similarly to what observed for the Fermi energy -- whose internally-consistent prediction is used as the point at which the DOS is computed. 

The prediction errors of the direct model of the band energy are low in comparison with the intrinsic variability, below $1\%$ and largely independent on the value of $\gb$. The fact that the error in the predicted band energy is largely independent on the smearing suggests that the averaging procedure that is associated with the evaluation of $\Eband$ reduces the sensitivity to the fine details of the DOS, and that the large error that is observed on the prediction of $\DOS(E)$ for $\gb=0.1$~eV is not reflected in coarser-grained features of the distribution of energy levels. 
The prediction errors of the indirect models are slightly lower than those of the direct model, once again with the exception of the PC decomposition, where the errors jump to $2\%$ for $\gb=0.5$~eV to $3\%$ for $\gb=0.1$~eV. Even though the use of PCs does help improve marginally the accuracy of the predicted DOS, this is clearly not reflected in the accuracy of derived quantities, because of the presence of high-frequency components in the DOS that contribute to the value of $\Efermi$, $\Eband$ and $\DOS(\Efermi)$ and are either discarded or very difficult to learn. 
Finally, the direct prediction errors of $A(\Delta)$ are usually low (between $2\%$ and $4.5\%$), with errors that grow gently as $\gb$ is reduced. The errors of the indirect models are systematically lower than the direct model, with the CDF model showing consistently the best performance. 

Overall, these examples show that using a model of the DOS as an intermediate step in the calculation of electronic-structure properties can out-perform, marginally or substantially, a direct prediction. The improvement is most noticeable when learning properties such as the excitation density $A(\Delta)$ or the DOS at the Fermi level that clearly depend in a non-trivial way on non-locality -- i.e. the presence of a localized defect can change in a non-additive manner the value of the property for the entire system. 
Another advantage of an indirect model is that, based on a single prediction, one can compute a multitude of physical observables -- in addition to those we mention here, the electronic contributions to a material's heat capacity or Gibbs free energy, the band gap, etc. -- and that these predictions are consistent with each other, rather than affected by independent model errors.
Contrary to what we observed when building a model of $\DOS(E)$, the strategy used to represent the target function has an important effect on the accuracy of the indirect predictions. In particular, learning the separate principal components in the data set leads consistently to degraded performance, in part because of the error incurred by truncating the PC expansion, but also in part because higher-order components have very poor learning rates, at least when using a single set of features for all the components. 
The CDF-based model is consistently the best of the indirect models, suggesting that a Wasserstein-type metric is the most relevant way to assess the quality of a predicted DOS. 
It is worth noting that the three approaches are simple linear transformations of the same data, which in a multivariate kernel regression framework is equivalent to the choice of a non-diagonal regularization of the regression weights that couples different targets properties. One could envisage to explicitly optimize the regularization to improve the accuracy in the desired derived quantities.

\begin{figure}
	\centering
	\includegraphics[width=1.\linewidth]{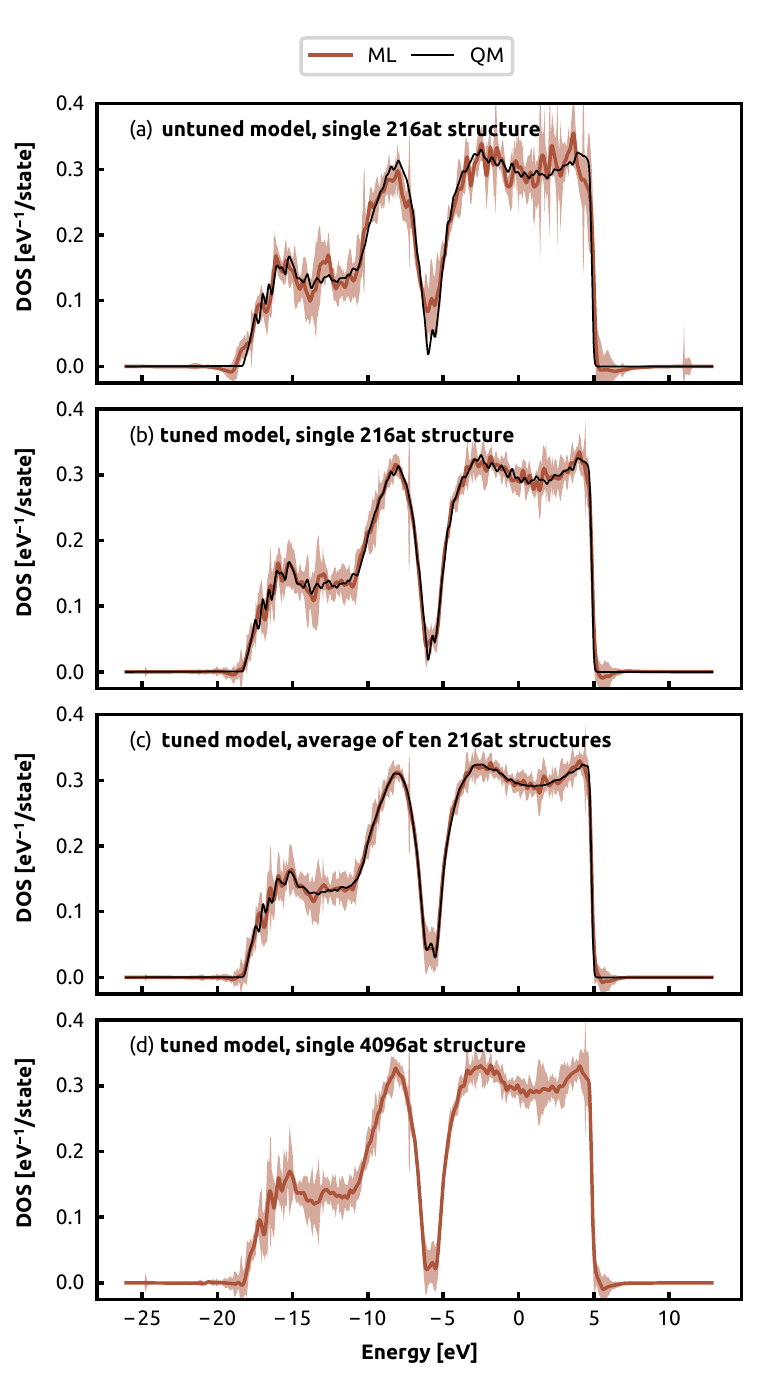}
	\caption{\revision{Comparison of DFT and ML electronic DOS of (a) a 216-atom structure where the ML DOS is based on an the general-purpose ML model discussed in Section~\ref{sec:results}, (b) a 216-atom structure where the ML DOS is based on a ML model that has been further tuned for this class of structures (c) an average of ten 216-atom structures with the tuned ML model (d) a 4096-atom structure with the tuned ML model.
	For all panels, the reference DOS is constructed using $\gb=0.1$~eV, and ML models use the CDF representation. The shaded area indicates the uncertainty in the ML prediction.}}
\label{fig:aSi_dos}
\end{figure}

\section{Density of states of amorphous silicon}\label{sec:a-si}

We use carefully-equilibrated large-scale configurations of amorphous silicon~\cite{Deringer2018} to demonstrate two advantageous features of a local machine-learning model, such as the one we use here. On the one hand, it allows predicting properties of large structures with a cost that scales linearly with system size; on the other, it provides a data-driven decomposition of the DOS in local contributions, that can be used to analyze structure-property relationships.   

\subsection{Large-scale evaluation of the DOS}

We consider a series of larger amorphous silicon (a-Si) structures, with a size ranging between 216 and 4096 atoms, that were obtained by slowly annealing a molten Si configuration using a GAP model, following the protocol described in Ref.~\cite{Deringer2018}. For all but the largest size, we compute DFT reference values following the same scheme we used for the train set.
\revision{As can be seen in Fig.~\ref{fig:kpca}, the larger size and careful equilibration leads to structures that are quite different from the 64-atoms a-Si, but very similar to each other, concentrated in a narrow region close to that occupied by liquid configurations. 
As shown in Fig.~\ref{fig:aSi_dos}, the general-purpose model we benchmark in the previous section achieves an accuracy comparable to that we observe for liquid configurations -- exhibiting clearly the qualitative features of the reference DOS. 
However, as noted earlier, a more finely-tuned training set is needed to achieve quantitative prediction accuracy with a high-resolution, $\gb=0.1$~eV DOS reference. To demonstrate that this fine-tuning can be easily achieved when focusing on a targeted application, we modify the training set by eliminating the cluster configurations, that occupy a completely disconnected portion of phase space, exhibit very large variance, and are built using a different band alignment reference with respect to bulk structures (see Fig~\ref{fig:kpca} and the SI). We also add 10 amorphous structures of 128 atoms each, \revision{generated by the same potential as the other large structures}, to ensure that the train set contains disordered configurations that are more representative of the large, slowly quenched configurations.} 
We represent the DOS using the CDF approach and with a Gaussian smearing $\gb=0.1$~eV, and use the same SOAP parameters we adopted in the previous section.

When computing the properties of materials in realistic thermodynamic conditions, it is necessary to average over multiple configurations, to compute a mean value that is consistent with a (quasi)-equilibrium probability distribution.
We observe that this ensemble average smooths the DOS, and that as a result the agreement between ML predictions and DFT reference values improves substantially (Fig.~\ref{fig:aSi_dos}).
The features of the DOS are well reproduced, including the presence of a small peak around the Fermi energy (i.e near $\approx -5.7$eV).
\revision{The cost of a ML prediction of the DOS is several orders of magnitude smaller than that of a DFT calculation even for the smaller system sizes. Furthermore, the cost of a GPR prediction scales linearly with system size as opposed to the cubic scaling of DFT (representative timings for a-Si samples of increasing size are reported in the SI). }
The ML model allows computing inexpensively the DOS of the largest structure (bottom right panel of Fig.~\ref{fig:aSi_dos}), for which an explicit DFT calculation would require application of a linear-scaling approach at still substantial cost\footnote{Cubic-scaling DFT simulations using the same setup we applied to smaller structures would require more the 1 million CPU hours and 20 TB of RAM.}. The DOS predicted for this structure is consistent with the average DOS of the smaller structures -- which indicates that this larger sample contains motifs that are found, with similar probability, in smaller simulations.

\begin{figure*}
	\centering
\includegraphics[width=1.0\linewidth]{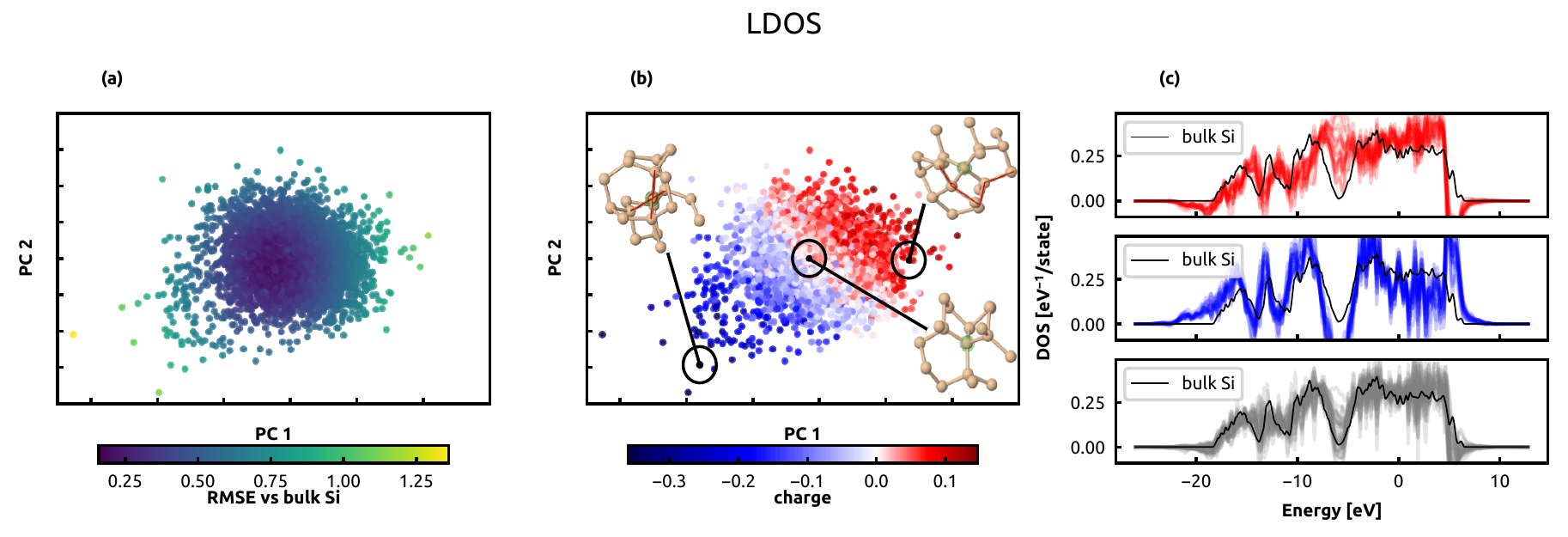}
	\caption{	\label{fig:MAP-LDOS}
		\revision{
            KPCovR map\cite{helf+20mlst} of the Si environments in the 4096-atom amorphous configuration, built based on a combination of SOAP kernels and LDOS predictions, with a mixing parameter $\alpha=0.05$. Points are colored according to (a) RMSE of the LDOS and (b) the local charge computed based on the LDOS. Snapshots of selected environments are also shown, with highly-distorted Si-Si-Si angles highlighted in dark red.  (c)  Comparison between the $\LDOS$ of selected configurations compared to the DOS of bulk Si. From top to bottom, the panels correspond to P, N, O type environments (respectively the 10 environments with the lowest, highest and median local charge). }}
\end{figure*}

\subsection{\revision{Interpreting the local ML DOS}}

\revision{
Having shown that the ML model reproduces accurately the total DOS of the sample, we assess whether the local contributions can be given a meaningful interpretation. 
The atom-centered decomposition of the DOS that underlies our model can yield LDOS contributions that are negative over some energy ranges. These negative contributions, that might appear unphysical, are a consequence of the fact that each atom-centered term reflects information from all of the atoms within the cutoff distance, so that atoms with large positive and negative $\LDOS(\Efermi)$ combine to yield the observed total DOS, which is the only physical observable given as target. 
As discussed in Refs.~\citenum{wilk+19pnas,veit+20jcp}, in the absence of an explicit, physics-based local learning target, atomic ML predictions reflect the interplay between structures and properties \emph{mediated by the choice of representation}. This is particularly relevant in light of the recent observation that 3-body correlation features are incomplete~\cite{pozd+20prl}, and that learning of global properties relies on neighboring atoms to disambiguate pairs of environments that have different structures but the same features (and hence the same ML-predicted local properties).

\begin{figure}[tbhp]
\centering
\includegraphics[width=1.0\linewidth]{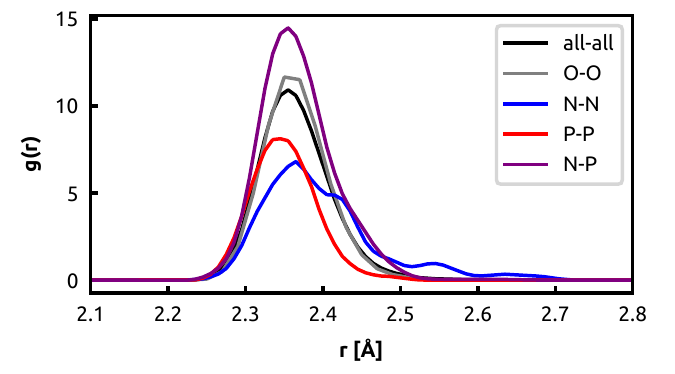}
\caption{
Pair correlation functions between Si atoms, resolved according to the classification between N (negatively-charged, $Q<-0.05$), P (positively charged, $Q>0.05$), O (neutral, $-0.05\le Q\le0.05$) atoms, introduced in the text.
	}
\label{fig:NOP-gr}
\end{figure}

\begin{figure*}[tbhp]
	\centering
\includegraphics[width=1.0\linewidth]{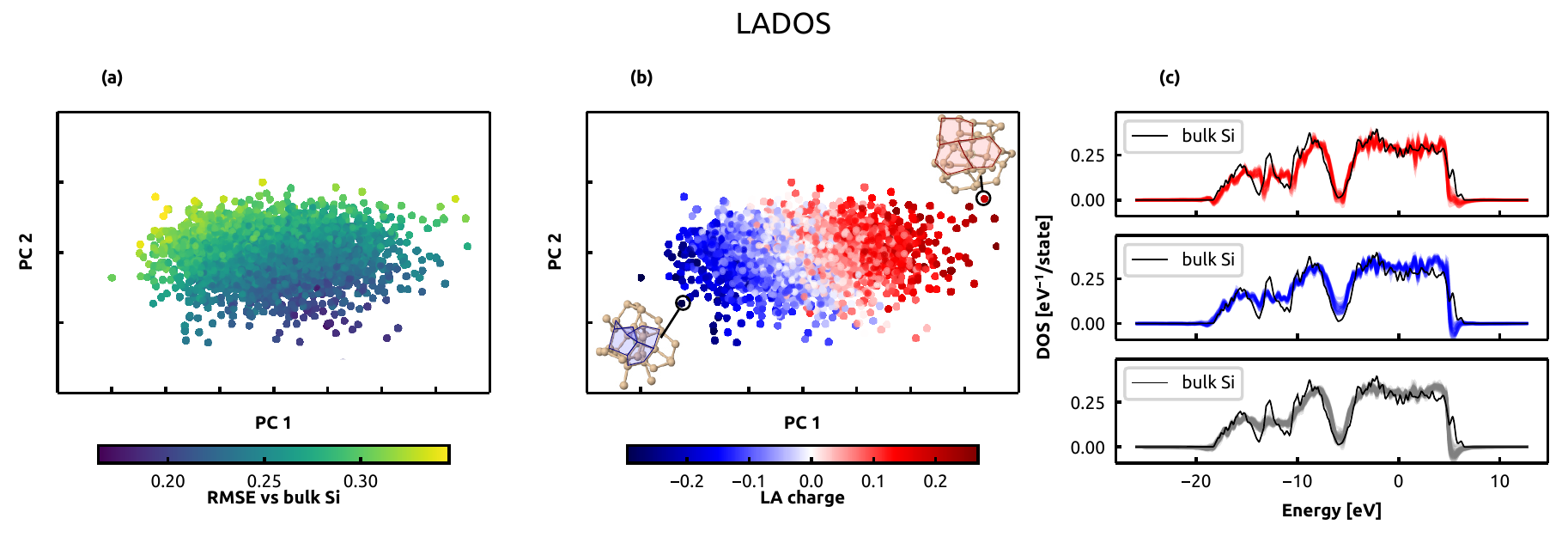}
	\caption{	\label{fig:MAP-LADOS}(left)
	\revision{
KPCovR map\cite{helf+20mlst} of the Si environments in the 4096-atom amorphous configuration, built based on a combination of SOAP kernels and LADOS predictions, with a mixing parameter $\alpha=0.05$. Points are colored according to (a) RMSE of the LADOS and (b) the local charge computed based on the LADOS. \revision{Snapshots of selected environments are also shown, with 5-rings highlighted in blue, and 7-rings highlighted in red.} (c)  Comparison between the $\LDOS$ of selected configurations compared to the DOS of bulk Si. From top to bottom, the panels correspond to P, N, O type environments (respectively the 10 environments with the lowest, highest and median local charge).}   }
\end{figure*} 

To investigate how the ML predictions of the LDOS can be used to analyze the structural motifs found in the 4096-atom a-Si structure,
we use the recently-developed kernel principal covariates regression technique (KPCovR)~\cite{kpcovr} to obtain a visually-interpretable description of the structure-property relations. 
KPCovR finds a low-dimensional projection of the kernel-induced features that correlate linearly with a set of target properties -- in this case the energy-resolved LDOS. 
As shown in Fig.~\ref{fig:MAP-LDOS}a, KPCovR constructs a purely structure-based latent space that correlates well with the ML LDOS, and identifies environments that have a DOS similar to bulk Si, and defective environments with substantially different electronic properties. 
We can further analyze the link between geometric and electronic structure by computing a ``local charge'' indicator, defined as 
\begin{equation}
Q(A_i) = 4-\int \operatorname{d}E\ f_{FD}(E-\Efermi) \LDOS(A_i,E).\label{eq:local-charge}
\end{equation}
This local charge correlates strongly with the KPCovR map (see Fig.~\ref{fig:MAP-LDOS}b). Type N environments (to the left of the plot) are negatively charged, type P environments (to the right of the plot) have a net positive local charge, and type O environments (in the middle, having a DOS most similar to bulk Si structures) are approximately neutral. 
By visualizing the environments with the interactive structure analyzer Chemiscope~\cite{frau+20joss,chemiscope} (input available in the SI), we can recognize the structural features associated with the principal axes of the KPCovR latent space. Type N environments have a distorted structure, with tetrahedral angles that approach 180 degrees. Type P environments have relatively regular distribution of nearest neighbors, but \emph{their neighbors} have a highly distorted configuration, similar to that observed for environments at the right end of the map. Environments at the center of the plot (type O) have both the central atom and its neighbors in a regular tetrahedral structure, and often with the same relative orientation one would find in crystalline Si structures. 
Looking more closely at the LDOS associated with these structures (Fig.~\ref{fig:MAP-LDOS}c) one sees that, as anticipated above, the atom-centered ML predictions exhibit strongly unphysical features, with large negative values of the LDOS being associated with P environments. 
The fact that physical values of the total DOS can arise by combining unphysical predictions is also apparent in the fact that the positions of N and P environments are strongly correlated. As shown in Fig.~\ref{fig:NOP-gr}, N and P atoms are less often found as first-neighbors of an environment of the same type, while N--P pairs are encountered more often than one would expect from a random distribution.  

One possibility to recover a more physically-interpretable prediction is to compute \emph{local averages} of the ML LDOS. In other terms, since each atom appears in multiple environments, it should get a share of the prediction for each of the environments it contributes to. 
This suggests the following expression for a locally-averaged DOS (LADOS):
\begin{equation}
\mathrm{LADOS}(E, A_i) = \sum_{j \in A} \frac{f_{\mathrm{cut}}(r_{ij})u(r_{ij}) \mathrm{LDOS}(E, A_j)}{\sum_{k \in A} f_{\mathrm{cut}}(r_{jk})u(r_{jk})},
\end{equation}
in which we use a weighting of the contributions that corresponds to that used to construct the local density features. This formulation also ensures that the same prediction for the global DOS of the structure can be obtained by summing over the LADOS values, as well as over the raw LDOS.
As shown in Fig.~\ref{fig:MAP-LADOS}, local averaging reduces the variability in the atom-centered predictions, and leads to largely positive-definite values of the local density of states -- which is consistent with the fact that averages over large portions of a structure tend to a well-defined total DOS. The principal axis of the KPCovR map is still largely correlated to a local charge value $Q(A_i)$ computed based on the LADOS, but the structural features that are associated with the local charge are less apparent, and more delocalized, than those we found for the raw ML LDOS. We observe that structures with a large positive local charge tend to be associated with rings of 7 Si atoms, while structures with a large negative charge tend to be close to many 5-membered rings. 
Furthermore, and the LDOS and LADOS-based values of $Q(A_i)$ correlate poorly with each other, and with commonly-used structural descriptors, such as the tetrahedrality index~\cite{CHAU1998, Errington2001} (for an interactive view, see the Chemiscope input in the SI), and do not show strong signals for the over- and under-coordinated structures that are found to exhibit a localized  band gap state in physics-based local DOS calculations\cite{Bernstein:2019dn}. 
An analysis of the atom-centered ML DOS, in combination with a hybrid supervised/unsupervised learning method such as KPCovR, facilitates the identification of the impact of structural heterogeneity on the electronic structure of disordered materials, but one should not over-interpret an analysis that is also influenced by the details of the ML representation and the regression scheme.

}
\section{\label{sec:conclusion}Conclusion}

In this work we present a ML framework based on sparse Gaussian process regression, a SOAP-based representation of local environment, and an additive decomposition of the electronic density of states to learn and predict the DFT-computed DOS for a diverse data set of Silicon structures, covering a broad range of thermodynamic conditions and different phases. We discuss the effect of the Gaussian broadening values usually used to smooth the DOS curves on the prediction process. We find that the large variance introduced by low smearing values contributes to increasing the validation error, because of the sharp peaks of the states. Errors grow from $8\%$ for a $0.5$eV smearing to $22\%$ for a $0.1$eV smearing. We also compare 3 different methods to represent the DOS, that can be linked to the use of different metrics to assess the error in the predictions: the pointwise discretization approach, a decomposition on the basis of selected principal components, and the description of the DOS as a derivative of its associated cumulative distribution function. We find that the different representations are fairly compatible, with a slight advantage to the latter for all the smearing values. 

We also investigate the accuracy of derived properties, that can be computed from the predicted DOS, against a direct ML model. We consider the Fermi energy, the DOS value at the Fermi energy, the band energy and the excitation spectrum. 
With the exception of the PC decomposition -- that gives poor results for the indirect prediction of $\Efermi$ and the band energy, partly due to errors introduced by the truncation of the principal components expansion -- we find that the indirect models lead to small but consistent improvements over direct predictions. 
This improvement is remarkable, because a direct model has the possibility to focus on the structure-property relations that are more relevant to the target. The fact that going through the DOS improves predictions indicates that the density of states is more amenable to an additive, local decomposition with respect to properties like $\Efermi$ that depend on the global imposition of charge neutrality.

We demonstrate an application of our ML model to the prediction of the DOS of some amorphous silicon configurations, including one containing 4096 atoms for which a brute-force DFT calculation would be prohibitively expensive. We observe excellent accuracy in the predictions, and that the averaging over multiple configurations -- that is necessary to obtain predictions consistent with experimental observations -- reduces considerably the discrepancy between the ML model and the DFT reference. 
A data-driven analysis of the local density of states reveals the interplay between structural motifs and electronic structure in a-Si, but a physical interpretation of the local DOS contributions computed by a ML model should not disregard the role played by the choice of structural features and regression scheme, which affects the atom-centered predictions in the absence of explicit reference values for the LDOS.

Our ML framework makes it possible to estimate, based exclusively on atomic configurations, one of the most essential descriptors of electronic structure. Combining it with one of the well-established potential energy models, this makes it possible to compute the electronic contributions to macroscopic properties such as the heat capacity of metals, to perform simulations that take into account finite-electronic-temperature effects~\cite{alav+94prl}, and provides another brick in the construction of a full surrogate ML model of the properties of molecules and materials.
The possibility of computing atomic charges by enforcing global charge neutrality, and then using local DOS to determine charge partitioning, provides an interesting line of investigation to realize a ``grand-canonical machine learning'' framework that combines a local model with a physics-based charge equilibration scheme.

\begin{acknowledgments}
The authors would like to thank Mariana Rossi for insightful discussion on band alignment in FHI-aims, and Rose Cersonsky for help performing KPCovR calculations. We thank V. L. Deringer for supplying a-Si configurations, and stimulating discussion on the definition of a locally-averaged DOS. MC and AA acknowledge funding from the European Research Council under the European Union's Horizon 2020 research and innovation programme (Grant Agreement No. 677013-HBMAP). CB acknowledges support by the Swiss National Science Foundation (Project No. 200021-182057).
\end{acknowledgments}

\end{document}